\documentclass[prc,aps,float,superscriptaddress,twocolumn,nofootinbib]{revtex4}

\usepackage{graphicx}
\usepackage[dvips]{color}
\usepackage{colordvi}
\usepackage{bm}
\usepackage{amsmath}
\usepackage{amssymb}
\usepackage{verbatim}
\usepackage{multirow}
\usepackage{bigstrut}
\usepackage{array}
\usepackage{fix-cm}
\usepackage{comment}

\newcommand{\ad}[1]{a_{#1}}
\newcommand{\ac}[1]{a^{\dagger}_{#1}}

\newcommand{\bra}[1]{\langle #1 \vert}
\newcommand{\ket}[1]{\vert #1 \rangle}

\begin{document}


\title{{\it Ab-initio} approach to effective single-particle energies in doubly closed shell nuclei}

\author{T. Duguet}
\email{thomas.duguet@cea.fr} \affiliation{CEA-Saclay DSM/Irfu/SPhN, F-91191 Gif sur Yvette Cedex, France} \affiliation{National Superconducting Cyclotron Laboratory and Department of Physics and Astronomy, Michigan State University, East Lansing, MI 48824, USA}

\author{G. Hagen}
\email{hageng@ornl.gov} \affiliation{Physics Division, Oak Ridge National Laboratory, Oak Ridge, TN 37831, USA} \affiliation{Department of Physics and Astronomy, University of Tennessee, Knoxville, TN 37996, USA}

\date{\today}

\pacs{21.10.Re, 21.60.Ev, 71.15.Mb}

\keywords{Nucleon transfer, Effective single-particle energy, Ab-initio method, Energy density functional}

\begin{abstract}
The present work discusses, from an ab-initio standpoint, the definition, the meaning, and the usefulness of effective single-particle energies (ESPEs) in doubly closed shell nuclei. We perform coupled-cluster calculations to quantify to what extent selected closed-shell nuclei in the oxygen and calcium isotopic chains can effectively be mapped onto an effective independent-particle picture. To do so, we revisit in detail the notion of ESPEs in the context of strongly correlated many-nucleon systems and illustrate the necessity to extract ESPEs through the diagonalization of the centroid {\it matrix}, as originally argued by Baranger. For the purpose of illustration, we analyse the impact of correlations on observable one-nucleon separation energies and non-observable ESPEs in selected closed-shell oxygen and calcium isotopes. We then state and illustrate the non-observability of ESPEs. Similarly to spectroscopic factors, ESPEs can indeed be modified by a redefinition of inaccessible quantities while leaving actual observables unchanged. This leads to the absolute necessity to employ consistent structure and reaction models based on the same nuclear Hamiltonian to extract the shell structure in a meaningful fashion from experimental data. 

\end{abstract}

\maketitle

\section{Introduction}
\label{intro}

The concept of single-nucleon shells dates back to the early days of nuclear physics and constitutes the basic pillar of the nuclear shell model
~\cite{goeppertmayer49}. The independent-particle approximation provides a zeroth-order picture of the structure of nuclei on top of which correlations are added to provide a more realistic description. Based on such a rationale, the correlated shell model has been able to explain the occurrence of extraordinarily stable configurations for specific neutron and proton numbers, known as magic numbers. As a matter of fact, the universal character of such magic numbers over the nuclear chart remains an open question today~\cite{Sorlin08a}. Recently, the evolution of shell structure and the understanding of the neutron drip-line location in oxygen isotopes have received considerable experimental and theoretical attention~\cite{baumann07a,janssens09a,Otsuka:2009cs}, e.g. significant shell gaps have been identified in $^{22}$O and $^{24}$O leading to the interpretation of new magic shell closures at $N=14, 16$ in $Z=8$ nuclei.

Identifying the underlying mechanisms responsible for the occurrence or the disappearance of magic numbers in specific regions of the nuclear chart requires improvement on the traditional shell model by allowing for a more systematic and consistent inclusion of correlations. In particular, questions related to the impact of continuum degrees of freedom~\cite{Tsukiyama:2009hy,Michel:2010zq,Tsukiyama:2010dn} and of three-nucleon forces on the evolution of nuclear shells is a frontier driving low-energy nuclear physics research in connection with radioactive ion beam facilities~\cite{Otsuka:2009cs,Holt:2010yb,Holt:2011fj}.

Whether a certain nucleon number qualifies as a (new) magic number cannot be postulated a priori. Experimentally, several quantities, e.g. the excitation energy and the collective character of the first $2^{+}$ state in even-even isotopes, the size of the gap in the one-nucleon addition/removal spectrum, and the spectroscopic factors of associated low-lying states in odd-even neighbours need to be extracted in order to make such an assessment. Theoretically, the same quantities need to be computed while including all many-body correlations that could play a role in order to check whether the picture associated with a magic number eventually holds.

It can be useful in this context, for analysis and interpretation purposes, to extract a single-nucleon shell structure, i.e. a set of {\it effective} single-particle energies (ESPEs) associated with an underlying independent-particle-like picture the system is mapped on. However, immediate non-trivial questions arise that are at the heart of the present study
\begin{enumerate}
\item Can a single-nucleon shell structure be unambiguously defined in a system that is intrinsically correlated? In other words, can ESPEs be computed on the sole basis of outputs of the many-body Schroedinger equation and not as a result of an a priori given zeroth-order approximation picture?
\item Correspondingly, to which auxiliary independent-particle problem are ESPEs related, i.e. which one-body Hamiltonian are ESPEs the eigenvalues of?
\item To what extent do correlations impact the effective independent-particle picture provided by such ESPEs?
\item In which way are ESPEs related to underlying nuclear forces?
\item Given that an unambiguous definition of ESPEs exists, is the associated simplified picture needed and beneficial or potentially misleading? In particular, are ESPEs physically observable quantities?
\end{enumerate}

Several of the above questions have been answered long ago while others still necessitate further clarifications. The procedure to extract ESPEs unambiguously (cf. point 1 above) goes back to French and Baranger~\cite{french66a,baranger70a,Umeya:2006eh} and can be utilized to address points 2, 3, 4, and 5. Such a procedure defines ESPEs as {\it centroid} energies denoting barycenters of correlated total binding energy differences between the A-nucleon state the one-nucleon transfer is performed on and the complete set of eigenstates of the A+1 and A-1 systems. Eventually, centroid energies can be related~\cite{baranger70a,Dufour:1995em,Umeya:2006eh} to the monopole part~\cite{bansal64a,zuker69a} of underlying nuclear interactions, which effectively answers points 2 and 4 above.

In spite of the existence of an unambiguous procedure to compute ESPEs, difficulties exist that can lead to improper conclusions, e.g. conclusions based on an analysis whose model dependence has not been properly identified and stated. On the experimental side, extracting a centroid energy necessitates the identification of all many-body states with a given $J^{\pi}$ from both one-nucleon stripping {\it and} pick-up reactions, which is not often possible. This is particularly critical as one moves away from doubly closed shell nuclei.


Theoretically, various levels of model dependence arise in the computation of ESPEs. On the deepest level, it is essential to understand that ESPEs depend, contrary to true observables, on the resolution scale $\Lambda$ used to define and solve the nuclear many-body problem. As a result, changing $\Lambda$ through, e.g., a unitary transformation on Fock space, changes ESPEs while leaving actual observables invariant. In this sense ESPEs are similar to spectroscopic factors; i.e. they can be used as a $\Lambda$-dependent analysis tool but cannot be seen as fundamental observable quantities. Moreover, and on a less fundamental level, approximations are often introduced in the computation of ESPEs that generate an artificial dependence on the single-particle basis used.
These various points will be discussed and illustrated in the present paper.

Difficulties may also arise when comparing ESPEs computed from an ab-initio approach on the one hand and from more effective methods, e.g. shell model and energy density functional, on the other. For instance, while the empirical shell-model "anchors" ESPEs on one-nucleon addition (removal) energies to (from) the closed-shell core nucleus of reference, this is not the case in an ab-initio context, as will be illustrated below.

The present paper follows the approach by Baranger as a way to delve further into the meaning and the usefulness of ESPEs by addressing questions 3 and 5 above, as well as by quantifying the error made when using approximations to Baranger's definition. The paper is organized as follows. Section~\ref{ESPE} collects essentially known results regarding the definition and the computation of ESPEs. Such a rather exhaustive introductory part is needed to discuss points that have often been overlooked over the years. Section~\ref{dysonequation} details the computation of ESPEs within the frame of the coupled-cluster (CC) method. Section~\ref{results} reports our results and illustrates various key properties of ESPEs. Specifically, the effect of correlations on both one-nucleon separation energies and on ESPEs is discussed, focusing first on a few specific examples before addressing systematics in oxygen and calcium isotopes. Starting from the ab-initio perspective provided by our results, the textbook rationale behind the truncated shell model is then briefly justified. Next, errors made by computing ESPEs in approximate ways are addressed before illustrating the deeper model dependence of ESPEs associated with their intrinsic resolution scale dependence. 
Conclusions are given in Sec.~\ref{conclusions}.

\section{Effective single-particle energies}
\label{ESPE}

In low-energy nuclear structure theory, one usually starts from an independent-particle model to convey the basic notions of single-particle states and shell structure. In this context, one resorts to systems that can be postulated {\it a priori} as being little influenced by correlations such that an effective independent-particle picture can be safely used. In a second step, actual correlations are introduced to explain, e.g., the fragmentation of the single-particle strength visible in one-nucleon transfer reactions. Such a pedagogical presentation makes it difficult to picture the possibility to define and extract {\it a posteriori} an {\it effective}, underlying single-particle shell structure in the presence of correlations, i.e. for A-body systems that are, strictly speaking, always correlated. It is thus more instructive to start from a realistic picture of the nucleus, i.e. a rather strongly correlated system, and extract a posteriori an effective single-particle shell structure from which correlations are to a large extent, but not entirely, screened out~\cite{baranger70a}.

To do so, we introduce the nuclear Hamiltonian under the form\footnote{The complication associated with the self-bound character of the nucleus, i.e. the need to subtract the center-of-mass motion in order to deal with internal many-body states and eigen-energies~\cite{Hagen:2010gd}, is overlooked in the present paper. Dealing with this difficulty in actual calculations is mandatory but would unnecessarily complicate the analytical expressions presented here without modifying significantly the outcome.} $H= T+V^{\text{2N}}+V^{\text{3N}}+\ldots$, where $T$ denotes the kinetic energy operator while $V^{\text{BN}}$ corresponds to a B-body interaction. We limit ourselves to 2N and 3N interactions throughout the formal part of the paper and to 2N forces in actual applications. Studying the impact of 3N interactions and forces of higher rank is postponed to future works. Given $H$, eigenstates and eigenenergies of the A-nucleon system are obtained by solving
\begin{equation}
H\ket{\Psi^{\text{A}}_{\mu}}=E^{\text{A}}_{\mu}\ket{\Psi^{\text{A}}_{\mu}} \,\,\,, \label{schroedinger}
\end{equation}
where the symmetry quantum number denoting the particle number has been singled out. The label $\mu$ collects a principal quantum number $n_\mu$, total angular momentum $J_\mu$, the projection of the latter along the $z$ axis $M_\mu$, parity $\Pi_\mu$ and isospin projection along the $z$ axis $T_\mu$ of the many-body state of interest. Use of the Greek label $\kappa_\mu$ will be made to denote the subset of quantum numbers $\kappa_\mu\equiv(\Pi_\mu,J_\mu,T_\mu)$. Due to rotational invariance of the nuclear Hamiltonian, eigenenergies $E^{\text{A}}_{\mu} \equiv E^{\text{A}}_{n_\mu \kappa_\mu}$ are independent of $M_\mu$.

In the following, we consider a spherical single-particle basis $\{\ac {p}\}$ appropriate to discussing the {\it spherical} shell structure. Basis states are labelled by $p \equiv \{ n_p, \pi_p, j_p, m_p, \tau_p \} \equiv \{ n_p, m_p, \alpha_p \}$, where $n_p$ represents the principal quantum number, $\pi_p$ the parity, $j_p$ the total angular momentum, $m_p$ its projection along the $z$-axis, and $\tau_p$ the isospin projection along the same axis.

We also consider the direct-product basis $\{b^{\dagger}_{\vec{r} \sigma \tau}\}$, where $\vec{r}$ is the position vector, $\sigma$ the projection of the nucleon spin along the $z$ axis, and $\tau$ its isospin projection.

\subsection{Spectroscopic amplitudes}
\label{spectro_amplitudes}

The physical processes providing information on the single-particle shell structure are one-nucleon transfer reactions. Although the discussion can be carried out for the transfer on any initial~\cite{Umeya:2006eh}. many-body state, we restrict ourselves in the following to the transfer on the ground state $\ket{\Psi^{\text{A}}_{0}}$ of an even-even system, i.e. a $J^{\pi} = 0^+$ state. Furthermore, we consider this nucleus to be of {\it doubly closed-shell} character\footnote{Such a notion relates to the filling of shells in the uncorrelated, e.g. Hartree-Fock, picture.}.

In this context, let us introduce $U_{\mu}$ ($V_{\nu}$) as the amplitude to reach a specific eigenstate $\ket {\Psi^{\text{A+1}}_{\mu}}$ ($\ket {\Psi^{\text{A-1}}_{\nu}}$) of the A+1 (A-1) system by adding (removing) a nucleon in a specific single-particle state to (from) the ground state of the A-body system $\ket {\Psi^{\text{A}}_{0}}$. Such spectroscopic amplitudes can be defined through their representation in any given single-particle basis. In basis $\{\ac {p}\}$, they read
\begin{subequations}
\label{eq:defu}
\begin{eqnarray}
U^{p}_{\mu} &\equiv& \bra {\Psi^{\text{A+1}}_{\mu}} a^\dagger_p \ket {\Psi^{\text{A}}_{0}}^{\ast}  \, , \\
V^{p}_{\nu} &\equiv& \bra {\Psi^{\text{A-1}}_{\nu}} a_p \ket {\Psi^{\text{A}}_{0}}^{\ast}  ,
\end{eqnarray}
\end{subequations}
whereas their representation in basis $\{b^{\dagger}_{\vec{r} \sigma q}\}$ provides the associated wave functions or {\it overlap functions}
\begin{subequations}
\label{eq:defu2}
\begin{eqnarray}
U_{\mu}(\vec{r} \sigma \tau) &\equiv& \bra {\Psi^{\text{A+1}}_{\mu}} b^\dagger_{\vec{r} \sigma \tau} \ket {\Psi^{\text{A}}_{0}}^{\ast}  \, , \\
V_{\nu}(\vec{r} \sigma \tau) &\equiv& \bra {\Psi^{\text{A-1}}_{\nu}} b_{\vec{r} \sigma \tau} \ket {\Psi^{\text{A}}_{0}}^{\ast}  .
\end{eqnarray}
\end{subequations}
An important property regarding the asymptotic behaviour of overlap functions derives from their equation of motion given by~\cite{Dickhoff:2004xx}
\begin{equation}
\label{dyson}
\left[h^{\infty} + \Sigma(\omega)\right]_{\omega=E^+_\mu} U_{\mu} =  E^+_\mu \,U_{\mu} \, ,
\end{equation}
and similarly for $(V_{\nu}, E_{\nu}^{-})$, where (observable) one-nucleon separation energies are defined through
\begin{subequations}
\label{sepenergies}
\begin{eqnarray}
E_{\mu}^{+} &\equiv& E^{\text{A+1}}_{\mu} - E^{\text{A}}_{0} \,\,\, , \\
E_{\nu}^{-} &\equiv& E^{\text{A}}_{0} - E^{\text{A-1}}_{\nu} \,\,\, .
\end{eqnarray}
\end{subequations}
The energy-{\it dependent} potential $\Sigma(\omega)$ denotes the {\it dynamical} part of the irreducible self-energy~\cite{Dickhoff:2004xx} that naturally arises in self-consistent Green's-function theory and that is to be evaluated at the eigensolution $E^+_\mu$ in Eq.~(\ref{dyson}). The static field $h^{\infty}$ is defined in Eq.~(\ref{HFfield}) and contains both the kinetic energy and the energy-{\it independent} part of the one nucleon self-energy. One can show from Eq.~(\ref{dyson}) that the long-distance behaviour of the radial part of the overlap function is governed by the corresponding one-nucleon separation energy, e.g. for $E^+_\mu<0$
\begin{equation}
\label{eq:rho_psympt_decomp}
U_{\mu}(r \sigma \tau) \underset{r\rightarrow +\infty}{\longrightarrow} A^{+}_{\mu} \, \frac{e^{-\varsigma^+_\mu\,r}}{\varsigma^+_\mu\,r} \, \, ,
\end{equation}
where $A^+_{\mu}$ denotes the so-called asymptotic normalization coefficient (ANC) while the decay constant is given by $\varsigma^+_\mu\equiv (-2m E^+_\mu/\hbar^2)^{1/2}$, where $m$ is the nucleon mass\footnote{Subtracting the center-of-mass motion would lead to using the reduced mass of the added/removed nucleon.}. A similar result can, of course, be obtained for $V_{\nu}(r \sigma \tau)$ whose decay constant $\varsigma^-_\nu$ relates to $E^-_\nu$.

From spectroscopic amplitudes one defines addition $S_{\mu}^{+}$ and removal $S_{\nu}^{-}$ spectroscopic probability matrices associated with states $\ket {\Psi^{\text{A+1}}_{\mu}}$ and $\ket {\Psi^{\text{A-1}}_{\nu}}$, respectively. Their matrix elements read in basis $\{\ac {p}\}$
\begin{subequations}
\label{spectroproba}
\begin{eqnarray}
S_{\mu}^{+pq} &\equiv&  \bra {\Psi^{\text{A}}_{0}} a_p \ket {\Psi^{\text{A+1}}_{\mu}} \bra {\Psi^{\text{A+1}}_{\mu}} a^\dagger_q \ket {\Psi^{\text{A}}_{0}}  \\
&=&   U^{p}_{\mu} \, U^{q \, \ast}_{\mu}  \nonumber  \, \, \, , \\
S_{\nu}^{-pq} &\equiv& \bra {\Psi^{\text{A}}_{0}} a^\dagger_q \ket {\Psi^{\text{A-1}}_{\nu}} \bra {\Psi^{\text{A-1}}_{\nu}} a_p \ket {\Psi^{\text{A}}_{0}} \\
&=& V^{p \, \ast}_{\nu} \, V^{q}_{\nu} \nonumber  \, \, \, ,
\end{eqnarray}
\end{subequations}
such that their diagonal parts, when expressed in the coordinate space basis, are nothing but {\it transition densities} for the one-nucleon transfer from $\ket {\Psi^{\text{A}}_{0}}$ to  $\ket {\Psi^{\text{A+1}}_{\mu}}$ and $\ket {\Psi^{\text{A-1}}_{\nu}}$, respectively.

Tracing the two spectroscopic probability matrices over the one-body Hilbert space ${\cal H}_1$ gives access to spectroscopic {\it factors}
\begin{subequations}
\label{spectrofactor}
\begin{eqnarray}
SF_{\mu}^{+} &\equiv&  \sum_{p \in {\cal H}_{1}} \left|U^{p}_{\mu}\right|^2 = \sum_{\sigma \tau} \int \!d\vec{r} \, \left|U_{\mu}(\vec{r} \sigma \tau)\right|^2 \, \, , \\
SF_{\nu}^{-} &\equiv& \sum_ {p \in {\cal H}_{1}} \left|V^{p}_{\nu}\right|^2 = \sum_{\sigma \tau} \int \!d\vec{r} \, \left|V_{\nu}(\vec{r} \sigma \tau)\right|^2 \,\, ,
\end{eqnarray}
\end{subequations}
which are nothing but the (basis-independent) norm of spectroscopic amplitudes. A spectroscopic factor characterizes to what extent an eigenstate of the A+1 (A-1) system can be described as a nucleon added to (removed from) a single-particle state on top of the ground-state of the A-nucleon system. Such a feature intrinsically depends on the resolution scale $\Lambda$ characterizing the nuclear Hamiltonian and is thus not, strictly speaking, observable~\cite{Furnstahl:2001xq,Jennings:2011jm}. Still, spectroscopic factors can serve as a tool to analyse the results obtained at a given resolution scale.

\subsection{Spectral function and strength distribution}

We now gather the complete spectroscopic information associated with one-nucleon addition and removal processes in the so-called spectral function $\mathbf{S}(\omega)\equiv \mathbf{S}^{+}(\omega) + \mathbf{S}^{-}(\omega)$. The spectral function denotes an energy-{\it dependent} matrix over ${\cal H}_1$ whose elements in basis $\{a^\dagger_p\}$ are defined through
\begin{eqnarray}
\mathbf{S}_{pq}(\omega) &\equiv& \!\!\!\!\! \sum_{\mu\in {\cal H}_{A\!+\!1}} \!\!\! S_{\mu}^{+pq} \,\, \delta(\omega -E_{\mu}^{+}) +  \!\!\!\!\!\sum_{\nu\in {\cal H}_{A\!-\!1}} \!\!\! S_{\nu}^{-pq}  \,\, \delta(\omega -E_{\nu}^{-}) , \nonumber
\end{eqnarray}
where the first (second) sum is restricted to eigenstates of $H$ in the Hilbert space ${\cal H}_{A\!+\!1}$ (${\cal H}_{A\!-\!1}$) associated with the A+1 (A-1) system. It is of interest to introduce the $n^{\text{th}}$ moment of the spectral function that defines an energy-{\it independent} matrix over ${\cal H}_1$ through
\begin{equation}
\mathbf{M}^{(n)} \equiv \int_{-\infty}^{+\infty} \omega^{n} \, \mathbf{S}(\omega) \, d\omega . \label{spec_funct_moments}
\end{equation}
One can easily obtain from $\{a_p,a^{\dagger}_q\} = \delta_{pq}$ that the zeroth-moment is nothing but the identity matrix
\begin{equation}
\mathbf{M}^{(0)}_{pq}  = \sum_{\mu\in {\cal H}_{A\!+\!1}} S_{\mu}^{+pq} + \sum_{\nu\in {\cal H}_{A\!-\!1}} S_{\nu}^{-pq} = \delta_{pq} \, , \label{normalizationspectro}
\end{equation}
such that the diagonal matrix element of $\mathbf{S}(\omega)$ possesses the meaning of a probability distribution function (PDF) in the statistical sense, i.e. the combined probability of adding and removing a nucleon to/from a specific single-particle basis state $| p \rangle$ integrates to 1 when summing over all final states in the A$\pm$1 systems.

Last, but not least, we introduce the spectral strength distribution (SDD) as the trace of the spectral function matrix
\begin{eqnarray}
{\cal S}(\omega) &\equiv& \text{Tr}_{{\cal H}_{1}}\left[\mathbf{S}(\omega)\right] \label{SDD_def}\\
&=& \!\!\!\!\! \sum_{\mu\in {\cal H}_{A\!+\!1}}\!\!\! SF_{\mu}^{+} \,\, \delta(\omega -E_{\mu}^{+}) + \!\!\!\!\! \sum_{\nu\in {\cal H}_{A\!-\!1}}  \!\!\! SF_{\nu}^{-}  \,\, \delta(\omega -E_{\nu}^{-}) \, , \nonumber
\end{eqnarray}
which is a basis-independent function of the energy.

\subsection{Independent-particle vs correlated systems}
\label{qualitativediscussion}

It is of pedagogical interest to discuss the typical patterns displayed by the spectral strength distribution for both independent-particle and correlated systems. The goal of this exercise is to illustrate in what sense observable one-nucleon separation energies cannot be interpreted as single-particle energies as soon as correlations are present in the system.

\begin{figure}
\includegraphics[scale=0.1]{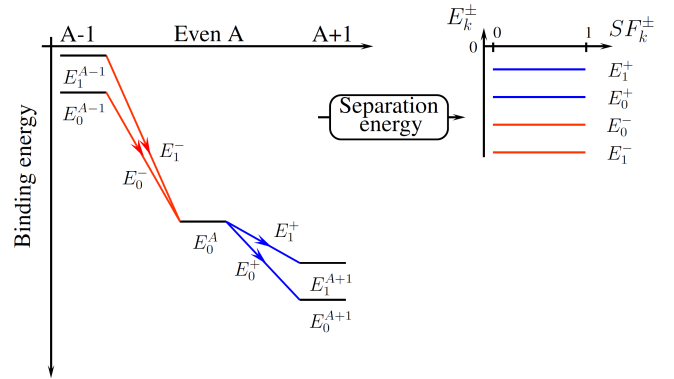}
\caption{(Color online) Schematic representation of one-nucleon addition and removal spectroscopic information for an independent-particle system. Left: binding energy for the ground-state of an even-even system and for the states of neighbouring nuclei reached by direct one-nucleon addition and removal processes. Right: corresponding spectral strength distribution.}
\label{SSD1}
\end{figure}

Figure~\ref{SSD1} provides a schematic display of one-nucleon addition and removal spectroscopic information for an independent-particle system. As many-body eigenstates of $H$ take the form of Slater determinants in such a case, there exists a particular single-particle basis of ${\cal H}_1$ in which addition and removal spectroscopic probability matrices read
\begin{subequations}
\label{SSDuncorrelated}
\begin{eqnarray}
S_{\mu}^{+pq}  &=& \delta_{p\mu} \, \delta_{pq} \, \delta_{pa} \, ,
\\
\nonumber
S_{\nu}^{-pq} &=& \delta_{p\nu} \, \delta_{pq} \, \delta_{pi} \, ,
\end{eqnarray}
\end{subequations}
where $i$ and $a$ characterize occupied ("hole") and unoccupied ("particle") states in the Slater determinant associated with the A-nucleon ground-state, respectively. Consequently, the many-body states reached by direct one-nucleon addition and removal processes are in one-to-one correspondence with single-particle basis states. As a result of such a bijection, one-nucleon separation energies are good candidates to play the role of single-particle energies. As a matter of fact, one has $E^{+}_\mu = \epsilon_a \, \delta_{a\mu}$ and $E^{-}_\nu = \epsilon_i \, \delta_{i\nu}$, where $\epsilon_a$ and $\epsilon_i$ denote eigenvalues of the one-body Hamiltonian governing the uncorrelated system associated with unoccupied and occupied single-particle states, respectively. Because the SDD integrates to the dimension of ${\cal H}_1$ by construction, spectroscopic factors of the corresponding states are equal to 1 whereas they are equal to zero for all the remaining states that are not reached by the direct one-nucleon transfer.

\begin{figure}
\includegraphics[scale=0.1]{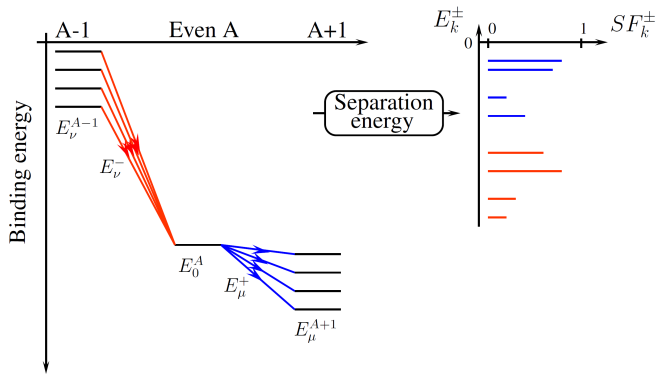}
\caption{(Color online) Same as Fig.~\ref{SSD1} for a correlated system.}
\label{SSD2}
\end{figure}

Let us now move to a correlated system. In {\it any} single-particle basis $\{a^{\dagger}_p\}$ of ${\cal H}_1$, $S_{\mu}^{+pq}$ ($S_{\nu}^{-pq}$) is now different from zero for any combination\footnote{Except for selection rules dictated by symmetries that lead, according to Eq.~(\ref{eq:defu3}), to $\pi_p=\pi_\mu$, $j_p=J_\mu$ and $\tau_p=T_\mu-T_0$.} of $\mu,p$ and $q$ ($\nu,p$ and $q$) indices. The SDD is thus fragmented as schematically displayed in Figure~\ref{SSD2}, i.e. a larger number of many-body states are reached through the direct addition and removal of a nucleon compared to the uncorrelated case\footnote{Of course, the dimension of ${\cal H}_{A\!+\!1}$ or ${\cal H}_{A\!-\!1}$ remains the same whether the system is correlated or not.}. Consequently, the number of peaks with non-zero strength in the SDD is greater than the dimension of ${\cal H}_1$, which forbids the establishment of a bijection between this set of peaks and any basis of  ${\cal H}_1$. Accordingly, and because the SDD still integrates to the dimension of ${\cal H}_1$ by construction (see Eq.~(\ref{normalizationspectro})), spectroscopic factors are smaller than one. The impossibility to realize such a bijection constitutes the most direct and intuitive way to understand why observable one-nucleon separation energies cannot be rigorously associated with single-particle energies when correlations are present in the system, i.e. as soon as many-body eigenstates of $H$ differ from Slater determinants.

\subsection{Effective single-particle energies}
\label{defESPE}

The discussion provided above underlines the fact that a rigorous definition of ESPEs is yet to be provided in the realistic context of correlated many-nucleon systems. A key question is: how can one extract a set of single-particle energy levels that (i) are in one-to-one correspondence with a basis of ${\cal H}_1$, (ii) are independent of the particular single-particle basis one is working with, (iii) are computable only using quantities coming out of the correlated A-body Schrodinger equation and that (iv)  reduce to HF single-particle energies in the HF approximation to the A-body problem. 

Let us make the hypothesis that ideal one-nucleon pick-up and stripping reactions have been performed such that separation energies $(E^{+}_\mu, E^{-}_\nu)$ and spectroscopic amplitudes (overlap functions) $(U_{\mu}(\vec{r} \sigma \tau), V_{\nu}(\vec{r} \sigma \tau))$ have been extracted consistently, i.e. in a way that is consistent with the chosen nuclear Hamiltonian $H(\Lambda)$ defined at a resolution scale $\Lambda$. In such a context, a meaningful definition of ESPEs does exist and goes back to French~\cite{french66a} and Baranger~\cite{baranger70a}. It involves the computation of the so-called centroid {\it matrix} which, in an arbitrary spherical basis of ${\cal H}_1$ $\{\ac {p}\}$, reads
\begin{subequations}
\begin{eqnarray}
h^{\text{cent}}_{pq} &\equiv& \sum_{\mu\in {\cal H}_{A\!+\!1}} S_{\mu}^{+pq} E_{\mu}^{+} + \sum_{\nu\in {\cal H}_{A\!-\!1}}  S_{\nu}^{-pq} E_{\nu}^{-} \label{defsumrule} \,\,\, ,
\end{eqnarray}
\end{subequations}
and is nothing but the first moment $\mathbf{M}^{(1)}$ of the spectral function matrix (see Eq.~\ref{spec_funct_moments}). Effective single-particle energies and associated states are extracted, respectively, as eigenvalues and eigenvectors of $h^{\text{cent}}$, i.e. by solving
\begin{eqnarray}
h^{\text{cent}} \, \psi^{\text{cent}}_p &=& e^{\text{cent}}_{p} \, \psi^{\text{cent}}_p \,\,\,, \label{HFfield3}
\end{eqnarray}
where the resulting spherical basis is denoted as $\{c^\dagger_{p}\}$. Written in that basis, centroid energies invoke diagonal spectroscopic probabilities\footnote{The definition of $e^{\text{cent}}_p$ sometimes incorporates the denominator $\sum_{\mu\in {\cal H}_{A\!+\!1}} S_{\mu}^{+pp} + \sum_{\nu\in {\cal H}_{A\!-\!1}} S_{\nu}^{-pp}$ in Eq.~(\ref{HFfield2}) to compensate for the possibility that, e.g. experimentally, normalization condition~\ref{normalizationspectro} might not be exhausted.}
\begin{eqnarray}
e^{\text{cent}}_{p} &\equiv&  \sum_{\mu\in {\cal H}_{A\!+\!1}} S_{\mu}^{+pp} E_{\mu}^{+} + \sum_{\nu\in {\cal H}_{A\!-\!1}}  S_{\nu}^{-pp} E_{\nu}^{-}  \,\,\,, \label{HFfield2}
\end{eqnarray}
and acquire the meaning of an average of one-nucleon separation energies weighted by the probability to reach the corresponding A+1 (A-1) eigenstates by adding (removing) a nucleon to (from) the single-particle state $\psi^{\text{cent}}_p$. Centroid energies are by construction in one-to-one correspondence with states of a single-particle basis of ${\cal H}_1$ which, as already pointed out before, is {\it not} the case of correlated one-nucleon separation energies with non-zero spectroscopic strength.

Equation~(\ref{HFfield3}) ensures that $\psi^{\text{cent}}_p(\vec{r}\sigma \tau)$ and $e^{\text{cent}}_{p}$ are consistent in the sense that the asymptotic behaviour of the former is driven by the latter, e.g. for $e^{\text{cent}}_{p}<0$ the radial part of the wave function behaves asymptotically as
\begin{equation}
\label{eq:rho_psympt_decomp2}
\psi^{\text{cent}}_p(r\sigma \tau) \underset{r\rightarrow +\infty}{\longrightarrow} C_{p} \, \frac{e^{-\xi_p\,r}}{\xi_p\,r} \, \, ,
\end{equation}
where $\xi_p\equiv(-2m e^{\text{cent}}_{p} /\hbar^2)^{1/2}$. Such a result underlines that single-particle wave-functions associated with ESPEs are intrinsically different from overlap functions $U_{\mu}(r \sigma \tau)$ ($V_{\nu}(r \sigma \tau)$) which are probed in transfer experiments. 

Experimentally, the extraction of ESPEs requires to collect the full spectroscopic strength up to high enough missing energies, i.e. the complete set of separation energies and cross sections from both one-nucleon stripping {\it and} pickup reactions. This unfortunately limits the possibility to perform sound comparisons on a systematic basis. Indeed, there are at best only a few nuclei along a given isotopic or isotonic chain that are characterized by complete enough spectroscopic data.

\subsection{Sum rule}
\label{defESPE2}

It is tedious but straightforward to prove that the n$^{\text{th}}$ moment of $\mathbf{S}(\omega)$ fulfils the identity
\begin{eqnarray}
\mathbf{M}^{(n)}_{pq}  &=& \langle \Psi^{\text{A}}_0| \{\overset{n \, \text{commutators}}{\overbrace{[\ldots[[\ad{p},H],H],\ldots]}},\ac{q}\} |\Psi^{\text{A}}_0\rangle \, . \label{identitymoments}
\end{eqnarray}
Using the second quantized form of $T$, $V^{\text{2N}}$, and $V^{\text{3N}}$, together with identities provided in Appendix~\ref{identities} and symmetries of interaction matrix elements, Eq.~(\ref{identitymoments}) applied to $n=1$ leads to~\cite{baranger70a,Polls94,Umeya:2006eh}
\begin{eqnarray}
h^{\text{cent}}_{pq} &=&  T_{pq} + \sum_{rs} \bar{V}^{\text{2N}}_{prqs} \, \rho^{[1]}_{sr} + \frac{1}{4}\sum_{rstv} \bar{V}^{\text{3N}}_{prtqsv} \, \rho^{[2]}_{svrt} \nonumber \\
&\equiv& h^{\infty} \,\,\, ,\label{HFfield}
\end{eqnarray}
where $\bar{V}^{\text{2N}}_{prqs}$ and $\bar{V}^{\text{3N}}_{prtqsv}$ are anti-symmetrized matrix elements and where
\begin{subequations}
\begin{eqnarray}
\rho^{[1]}_{pq} &\equiv&  \bra{\Psi^{\text{A}}_0} \ac{q} \ad{p} \ket{\Psi^{\text{A}}_0} = \sum_{\mu} {V^{p}_\mu}^{\ast} \, V^{q}_\mu \,\,\, , \\
\rho^{[2]}_{pqrs} &\equiv&  \bra{\Psi^{\text{A}}_0} \ac{r} \ac{s} \ad{q} \ad{p} \ket{\Psi^{\text{A}}_0}  \,\,\, ,
\end{eqnarray}
\end{subequations}
denote one- and two-body density matrices of the {\it correlated} A-body ground-state, respectively. The static field $h^{\infty}$, already introduced in Sec.~\ref{spectro_amplitudes}, contains both the kinetic energy and the energy-{\it independent} part of the one-nucleon self-energy in the A-body ground state~\cite{Polls94}.

\begin{figure}
\includegraphics[scale=0.13]{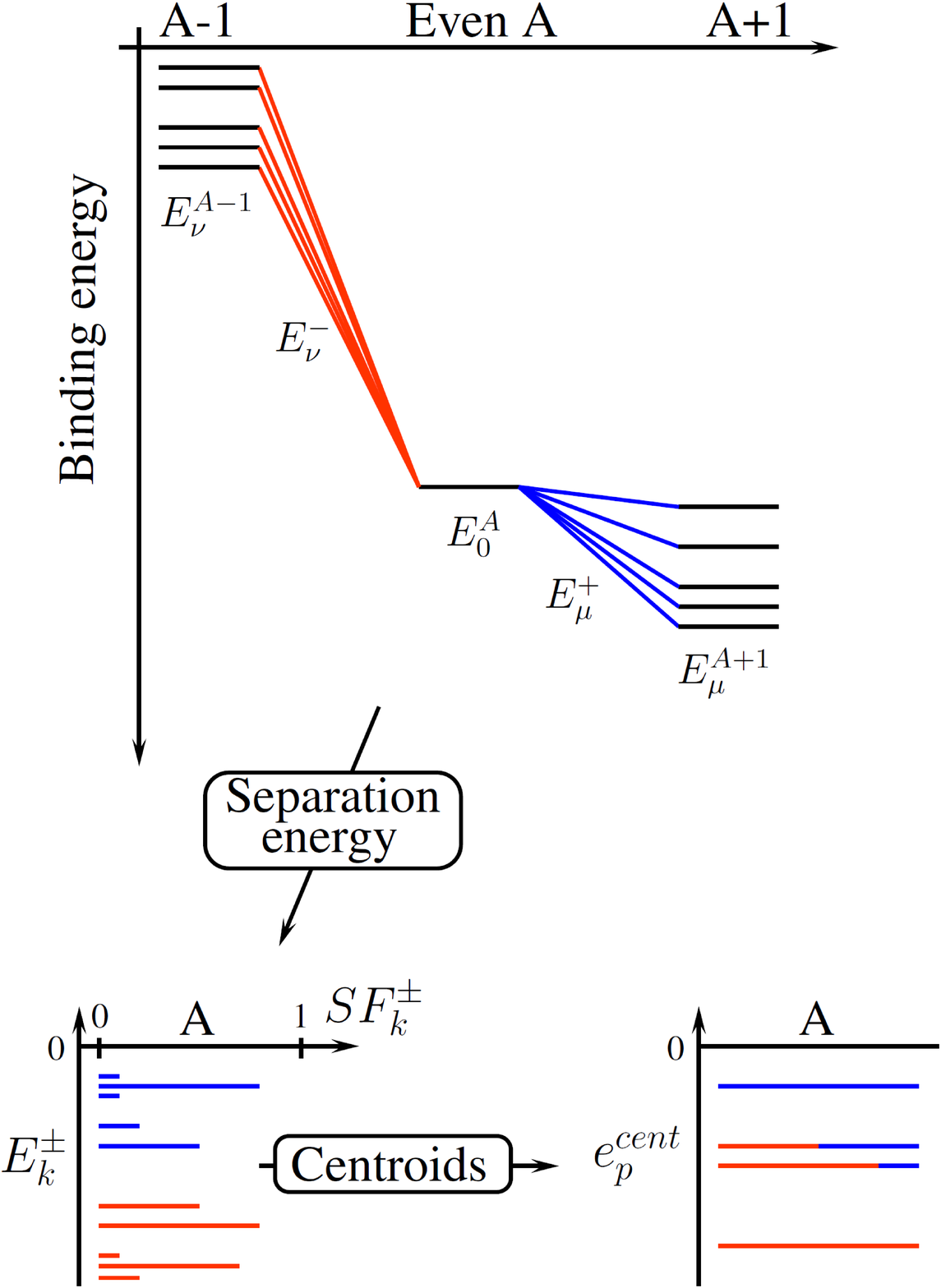}
\caption{(Color online) Schematic picture. Top: total binding energies (Eq.~\ref{schroedinger}) of three successive nuclei and associated one-nucleon addition/removal energies (Eq.~\ref{sepenergies}) from the ground state of the intermediate system. Bottom left: spectral strength distribution (Eq.~\ref{SDD_def}). Bottom right: corresponding ESPE spectrum (Eq.~\ref{HFfield2}). The color coding underlines that ESPEs close to the Fermi energy contain significant contributions from both addition and removal channels.}
\label{fig:stagg1}
\end{figure}

Equation~(\ref{HFfield}) demonstrates that the centroid matrix is a one-body field possessing a simple structure and an intuitive meaning. In particular, the centroid field reduces to the Hartree-Fock (HF) mean field in the HF approximation. As a result, ESPEs are nothing but HF single-particle energies in such a case and are equal to one-nucleon separation energies according to Koopmans' theorem~\cite{koopmans34}. Consistently, overlap, centroid, and HF single-particle wave-functions coincide in that limit. Of course, centroid energies also reduce to eigenvalues of the one-body Hamiltonian in the limit of an uncorrelated system. When correlations beyond HF are switched on, ESPEs are modified through the presence of correlated density matrices in Eq.~(\ref{HFfield}); i.e. the B-nucleon interaction is folded with the correlated (B-1)-body density matrix  $\rho^{[\text{B-1}]}$. Through that transition, ESPEs continuously evolve as centroid energies rather than as observable separation energies such that Koopmans' theorem does not hold any more. Centroid energies are schematically compared to observable binding and separation energies in Figure~\ref{fig:stagg1}.

On the practical side, Eq.~(\ref{HFfield}) underlines that the averaged information contained in ESPEs only requires the computation of the A-body ground-state. As long as one is not interested in the full spectroscopic strength of the A$\pm$1 systems but only in their centroids, one only needs to compute one nucleus instead of three. In practice however, Eq.~(\ref{eq:efinal}) is rarely computed in terms of the correlated density matrix, e.g. shell-model applications usually invoke a filling approximation typical of an independent-particle approximation. This is believed to be a decent approximation as long as (i) low-lying states carry a major part of the single-particle spectroscopic strength, as for the transfer on a doubly closed-shell nucleus, and (ii) nucleons of the other species are themselves not strongly correlated, because of pairing for example. See, e.g., Ref.~\cite{smirnova04a} and references therein for a related discussion. Such an issue becomes critical whenever one is looking into, e.g., the neutron shell structure of a neutron open-shell nucleus. In such a situation, a normal filling is inappropriate and it is mandatory to fold the monopole interaction in Eq.~(\ref{eq:efinal}) with a density matrix reflecting the presence of correlations in the system.

Using that the even-even ground state the one-nucleon transfer is performed on is a $J^{\Pi}=0^{+}$ state, Wigner-Eckart's theorem allows one to obtain the explicit dependence of spectroscopic amplitudes on $m_p$ and $M_\mu$, i.e.
\begin{subequations}
\label{eq:defu3}
\begin{eqnarray}
U^{p}_{\mu} &\equiv& U^{n_p \, [\alpha_p]}_{n_\mu}  \, \delta_{\kappa_\mu \alpha_p} \, \delta_{M_\mu m_p}  \, , \\
V^{p}_{\nu} &\equiv& V^{n_p \, [\alpha_p]}_{n_\nu} \, \delta_{\kappa_\nu \alpha_p} \, \delta_{M_\nu -m_p} \, (-1)^{m_p}  ,
\end{eqnarray}
\end{subequations}
such that the single-particle operator picks out the angular momentum, the parity and the isospin projection of the A$\pm$1 state the transfer goes to; i.e. $j_p=J_\mu$, $\pi_p=\Pi_\mu$ and $\tau_p=T_\mu-T_0$. Consequently, the one-body density matrix of the A-body ground state reads
\begin{equation}
\rho^{[1]}_{pq} \equiv \rho_{n_p n_q}^{[\alpha]} \, \delta_{\alpha_p \alpha_q} \, \delta_{m_p m_q} \,  , \label{OBDM}
\end{equation}
such that, retaining the 2N force only for simplicity and expressing its anti-symmetrized matrix elements in a jj-coupled scheme, 
one obtains~\cite{baranger70a,Dufour:1995em,Umeya:2006eh} in basis $\{c^\dagger_p\}$
\begin{equation}
\label{eq:efinal}
e^{\text{cent}}_{n_p[\alpha_p]} = t^{[\alpha_p]}_{n_p n_p} +  \sum_{n_q n_r} \sum_{\alpha_q} \bar{v}^{[\alpha_p \alpha_q \alpha_p \alpha_s]}_{n_p n_q n_p n_r}
\, \rho_{n_r n_q}^{[\alpha_q]}  \,\,\,,
\end{equation}
where $\bar{v}^{[\alpha_p \alpha_q \alpha_p \alpha_q]}_{n_p n_q n_p n_r}$
is the reduction of the 2N interaction to its so-called {\it monopole}, i.e. angular averaged, part. Higher multipoles and in particular the quadrupole part that drives the dominant part of correlations are screened out from ESPEs.

\subsection{Resolution-scale dependence}
\label{sec_scale_dependence}

Let us briefly explain the intrinsic resolution-scale dependence of ESPEs. Such a feature derives directly from the resolution scale of spectroscopic amplitudes~\cite{Furnstahl:2001xq,Jennings:2011jm} entering the definition of the centroid matrix, see Eq.~(\ref{defsumrule}). Following the philosophy of the similarity renormalization group (SRG)~\cite{Bogner:2006pc,Anderson:2010aq}, we consider a change of resolution scale via a unitary transformation $U(\Lambda)$ of the Hamiltonian
\begin{subequations}
\label{unitary1}
\begin{eqnarray}
H(\Lambda)&\equiv& U(\Lambda) \, H \, U^{\dagger}(\Lambda)  \label{unitary1a}\\
&\equiv& T + V^{\text{2N}}(\Lambda) + V^{\text{3N}}(\Lambda) + \ldots \: ,  \label{unitary1b}
\end{eqnarray}
\end{subequations}
where the scale characterizing the initial $H$ is omitted for simplicity. As can be trivially shown, Eq.~(\ref{unitary1a}) induces a transformation of eigenvectors of Eq.~(\ref{schroedinger})
\begin{eqnarray}
| \Psi^{\text{A}}_{\mu} (\Lambda) \rangle &\equiv& U(\Lambda) \, | \Psi^{\text{A}}_{\mu} \rangle \: , \label{unitary2}
\end{eqnarray}
such that\footnote{In practical applications, such an invariance is broken to some extent due to the approximate way of performing the transformation of the Hamiltonian, e.g. neglecting induced many-body interactions in Eq.~(\ref{unitary1b}), and to approximations performed when solving the A-body problem~\cite{Jurgenson:2009qs,Roth:2011ar}. However, the discussion of the present section is concerned with tracking what happens in the hypothesis of an exact unitary transformation and an exact solution of the A-body Schroedinger equation.} the associated observable, i.e. the eigenenergy, remains unchanged $E^{\text{A}}_{\mu}(\Lambda)=E^{\text{A}}_{\mu}$. Similarly, any observable associated with a Hermitian operator $O$ must remain invariant, which imposes the transformation of $O$ according to $O(\Lambda) \equiv U(\Lambda) \, O \, U^{\dagger}(\Lambda)$.

Let us now come to ESPEs. The key difference from an observable resides in the fact that the very nature of ESPEs is to inform us of effective single-nucleon degrees of freedom inside the nuclear medium, independently of the form of the Hamiltonian. In other words, the {\it choice} is made to keep the {\it definition} of ESPEs independent of $\Lambda$. Before or after transformation~\ref{unitary1a}, ESPEs are always extracted through Eq.~(\ref{defsumrule}), where $S^{\pm \, pq}_\mu(\Lambda)$ retains the same formal expression as before, i.e. they invoke spectroscopic amplitudes computed through
\begin{subequations}
\label{eq:defu_unitarytransformed}
\begin{eqnarray}
U^{p}_{\mu}(\Lambda) &\equiv& \bra {\Psi^{\text{A+1}}_{\mu}(\Lambda)} a^\dagger_p \ket {\Psi^{\text{A}}_{0}(\Lambda)}^{\ast}  \, , \\
V^{p}_{\nu}(\Lambda) &\equiv& \bra {\Psi^{\text{A-1}}_{\nu}(\Lambda)} a_p \ket {\Psi^{\text{A}}_{0}(\Lambda)}^{\ast}  .
\end{eqnarray}
\end{subequations}
Contrary to the many-body states involved, operators $a^\dagger_p$ and $a_p$ are {\it not} transformed in Eq.~(\ref{eq:defu_unitarytransformed}), which generates automatically an intrinsic dependence of $U^{p}_{\mu}$ and $V^{p}_{\nu}$ on $\Lambda$. One could, of course, choose to transform operators $a^\dagger_p$ and $a_p$ in the definition of spectroscopic amplitudes in order to make the latter invariant under the unitary transformation. However, transforming $a^\dagger_p$, e.g., would result into a linear combination of operators of the form $a^\dagger_q$, $a^\dagger_q a^\dagger_r a_s$, \ldots such that the spectroscopic amplitude would not provide the information one was after in the first place, e.g. the overlap between eigenstates of $H(\Lambda)$ in the A+1 (A-1) system and the state obtained by adding (removing) a nucleon to (from) a given single-particle state $| p \rangle$ on top of the A-body ground-state. Eventually, the resolution-scale dependence of spectroscopic amplitudes propagates to their norm, i.e. spectroscopic factors~\cite{Furnstahl:2001xq,Jennings:2011jm}, and to ESPEs.

The discussion provided above points to an important conclusion. The information one is sometimes after, e.g. computing spectroscopic factors and ESPEs, is not necessarily observable. Such an information is not absolute and can be modified by a redefinition of inaccessible quantities, i.e. the Hamiltonian and its eigenvectors in the present case, which leaves of course true observables untouched. It remains to be seen how much ESPEs are changed in actual calculations by varying the resolution scale $\Lambda$ over a reasonable interval of interest. This question is addressed in Sec.~\ref{results}. It could very well be that the induced variation of the ESPEs is negligible compared to other sources of uncertainties, e.g. approximations in their computation. Still, it is of prime importance to keep such an intrinsic model dependence of ESPEs in mind.

\section{Coupled cluster method}
\label{dysonequation}

One-nucleon separation energies and spectroscopic amplitudes introduced respectively in Eq.~(\ref{sepenergies}) and Eqs.~(\ref{eq:defu}-\ref{eq:defu2}) are defined without any reference to a particular method used to solve the many-body problem.

We are presently interested in using the ab-initio coupled-cluster method (CCM). Let us briefly outline the procedure to compute ground and excited states of a closed (sub-)shell nucleus A and of odd A$\pm$1 neighbours within CCM. From there, all needed quantities to compute ESPEs can be extracted. In CCM, the exact ground state is written in the exponential form
\begin{equation}
  \ket{\Psi^{\text{A}}_{0}} = e^{T}\ket{\Phi_0},
  \label{CC1}
\end{equation}
where $\ket{\Phi_0} $ is an uncorrelated single-reference Slater determinant
built from a convenient spherical single-particle basis, usually chosen
as mean-field HF orbitals. Many-body correlations beyond the
mean field are introduced by the operator $T = T_1 + T_2 + \ldots +T_A$,
which is a linear expansion in $n$-particle-$n$-hole excitation operators $T_n$, with $n=1,\ldots,A$.

The only approximation occurring in CCM regards the truncation
of $T$ to a given low-lying excitation level; e.g. $T \approx T_1 + T_2 $ is
the most commonly used approximation known as the coupled-cluster method
with single and double excitations (CCSD). Inserting the coupled-cluster ansatz
(\ref{CC1}) into the A-body Schr\"odinger equation (Eq.~(\ref{schroedinger}))
and projecting from the left with $ \bra{\Phi_0}e^{-T}, \bra{\Phi_i^a}e^{-T}$
and $\bra{\Phi_{ij}^{ab}}e^{-T}$ respectively, coupled-cluster equations are obtained under the form
\begin{subequations}
\label{cc2}
\begin{eqnarray}
\bra{\Phi_0}e^{-T} H e^{T} \ket{\Phi_0}&=&E^{\text{A}}_{0}, \\
\bra{\Phi_i^a}e^{-T} H e^{T} \ket{\Phi_0}&=&0, \\
\bra{\Phi_{ij}^{ab}}e^{-T} H e^{T} \ket{\Phi_0}&=&0.
\end{eqnarray}
\end{subequations}
These equations determine the unknown amplitudes entering $T_1$ and $T_2$ as well as the ground state energy $E^{\text{A}}_{0}$.
Here $ \bra{\Phi_i^a} $ and $\bra{\Phi_{ij}^{ab}}$ are one-particle-one-hole and
two-particle-two-hole excited reference states.

Equation~(\ref{cc2}) underlines that the similarity-transformed Hamiltonian
$\bar{H} \equiv e^{-T} H e^{T} $ plays a key role such that its ground state
is nothing but the reference state $\ket{\Phi_0} $. The operator $\bar{H}$ is not Hermitian, which implies that
coupled-cluster theory is manifestly non-variational. The non-variational nature
of CCM makes it necessary to access both right {\it and} left eigenstates
of $\bar{H}$ to compute associated one- and two-body density matrices. Such eigenstates of $\bar{H}$ can be computed through the so-called equation-of-motion coupled-cluster
method (EOM-CCM). The idea of EOM-CC is essentially to diagonalize $\bar{H}$ within
a subspace of $n$-particle-$m$-hole excited reference functions. Within the EOM-CCSD
approximation, right and left eigenstates of closed-shell nucleus A are
given by
\begin{subequations}
\begin{eqnarray}
\ket{R^A_\mu} =  R^A_\mu\ket{\Phi_0} \, \, , \\
\bra{L^A_\mu} = \bra{\Phi_0}L^A_\mu \, \, ,
\end{eqnarray}
\end{subequations}
where $R^A_\mu$ ($L^A_\mu$) is a linear combination of one-particle-one-hole and two-particle-two-hole (de-)excitation operators.
Similarly, EOM-CC is the method of choice to access eigenstates of odd A$\pm$1 neighbouring nuclei according to
$ \ket{R^{A\pm1}_\mu} =  R^{A\pm1}_\mu \ket{\Phi_0^{A}}$ and
$ \bra{L^{A\pm1}_\mu} = \bra{\tilde{\Phi}_0^{A}}L^{A\pm1}_\mu $, where now $R^{A\pm1}_\mu \, (L^{A\pm1}_\mu)$
denotes a linear combination of one-particle (one-hole) and
two-particle-one-hole (one-particle-two-hole) (de-)excitation operators
(see for example Ref.~\cite{Bartlett02} for further details).
Left and right eigenstates form a bi-orthogonal set, i.e.
\begin{equation}
\bra{L_\mu}R_{\mu'}\rangle  = \delta_{\mu\mu'}.
\end{equation}
where we have dropped the superscript referring to nucleus A and A$\pm$1.
Right eigenstates $R_\mu$ are solutions of the eigenvalue problem
\begin{align}
    (\bar{H} R_\mu)_C |\Phi_0\rangle  &= E_\mu R_\mu |\Phi_0\rangle \ ,
    \label{EOM_master}
\end{align}
and similarly for left eigenstates $L_\mu$. Here,
$(\bar{H} R_\mu)_C$ denotes all terms that connect $\bar{H}$ with $R_\mu$.
The one- and two-body density matrices of the A-body ground state
together with one-nucleon spectroscopic amplitudes and probabilities can now be computed according to
\begin{subequations}
\label{CC_dens}
\begin{eqnarray}
\rho^{[1]}_{pq} & \equiv & \bra{\Phi_0} L^A_0 \overline{\ac{q} \ad{p}} \ket{ \Phi_0} \, \, , \\
\nonumber
\rho^{[2]}_{pqrs} & \equiv &  \bra{\Phi_0} L^A_0 \overline{\ac{r} \ac{s} \ad{q} \ad{p}} \ket{ \Phi_0} \, \, ,
\end{eqnarray}
\end{subequations}
and to
\begin{subequations}
\label{CC_SFamp}
\begin{eqnarray}
S_{\mu}^{+pq} & \equiv &  \bra{\Phi_0}L^A_0 \overline{a_p} R^{A+1}_\mu\ket{\Phi_0} \bra{\Phi_0}L^{A+1}_\mu \overline{a_q^\dagger}\ket{\Phi_0}  \, \, , \\
S_{\mu}^{-pq} & \equiv & \bra{\Phi_0}L^A_0  \overline{a_q^\dagger} R^{A-1}_\mu\ket{\Phi_0} \bra{\Phi_0}L^{A-1}_\mu \overline{a_p} \ket{\Phi_0} \, \, .
\end{eqnarray}
\end{subequations}
Using the Baker-Campbell-Hausdorff commutator expansion, one can derive finite and closed-form algebraic expressions for similarity-transformed operators $\overline{a_p}$, $\overline{a_p^\dagger} $, $ \overline{\ac{q} \ad{p}} $ and $\overline{\ac{r} \ac{s} \ad{q} \ad{p}} $. See Refs.~\cite{Jensen2010,Jensen2011} for details on derivation and computation of spectroscopic factors within coupled-cluster theory.

\section{Results}
\label{results}

Results shown below have been obtained using a 2N force only,
thereby omitting forces of higher rank. In order to improve convergence properties and
make the nuclear many-body problem more perturbative, we use a soft $V_{\text{low-k}} $
2N interaction ~\cite{bogner03} obtained with a smooth regulator ~\cite{bogner07a} for
various cutoff values between $\Lambda=$2.0\,fm$^{-1}$ and 3.0\,fm$^{-1}$. These soft $V_{\text{low-k}} $ interactions
are obtained by evolving down the N$^{3}$LO Chiral interaction~\cite{entem03} with cutoff $\Lambda_{\chi}=$500\,MeV.
For the single-particle model space, we use the HF basis built from
N+1$=13$ major oscillator shells, with a fixed oscillator frequency of
$\hbar\omega$ =16\,MeV. This model space is sufficient to obtain fully converged results
for medium mass nuclei with soft $V_{\text{low-k}} $ interactions (see Ref.\cite{Hagen:2010gd}).


\subsection{Turning on correlations in a controlled way}
\label{switching}

Let us first illustrate in a pedagogical manner the effect of correlations on both one-nucleon separation energies and ESPEs. To do so, we apply Wick's theorem with respect to the HF vacuum $| \Phi^{\text{HF}}_0 \rangle$ and write the Hamiltonian in normal-ordered form using the HF single-particle basis $\{d^{\dagger}_p\}$, i.e.
\begin{eqnarray}
H &=& h^{\text{HF}} +V_{\rm res} \,\,\, ,
\end{eqnarray}
where
\begin{subequations}
\begin{eqnarray}
h^{\text{HF}}  &\equiv& E^{\text{HF}}_0   +  \sum_{p} \epsilon^{\text{HF}}_{p} \,  :d^{\dagger}_p d_p: \ \,\,\, , \\
V_{\rm res} &\equiv& \frac{1}{4} \sum_{pqrs} \bar{V}^{\text{2N}}_{pqrs} : d^{\dagger}_{p} d^{\dagger}_{q} d_{s} d_{r}:  \,\,\, ,
\end{eqnarray}
together with
\begin{eqnarray}
E^{\text{HF}}_0  &\equiv&  \sum_{p=1}^A T_{pp} + \frac{1}{2} \sum_{p,q=1}^A \bar{V}^{\text{2N}}_{pqpq}  \,\,\, ,\\
\epsilon^{\text{HF}}_{p}  &\equiv&  T_{pp} + \sum_{q=1}^A \bar{V}^{\text{2N}}_{pqpq}  \,\,\, .
\end{eqnarray}
\end{subequations}
Scaling the residual interaction $V_{\rm res}$ by a factor $\lambda \in [0,1]$, one defines a parameter-dependent Hamiltonian $H_{\lambda} \equiv h^{\text{HF}} + \lambda \, V_{\rm res}$ that tunes correlations between the two limits of interest, i.e. from the uncorrelated regime $H_{0}=h^{\text{HF}}$ to the fully correlated regime $H_{1} = H$. Eventually, we solve EOM-CC equations repeatedly for several values of $\lambda \in [0,1]$ and for the specific cutoff value $\Lambda=$2.4\,fm$^{-1}$.

\begin{figure}
\includegraphics[width=0.45\textwidth,clip=]{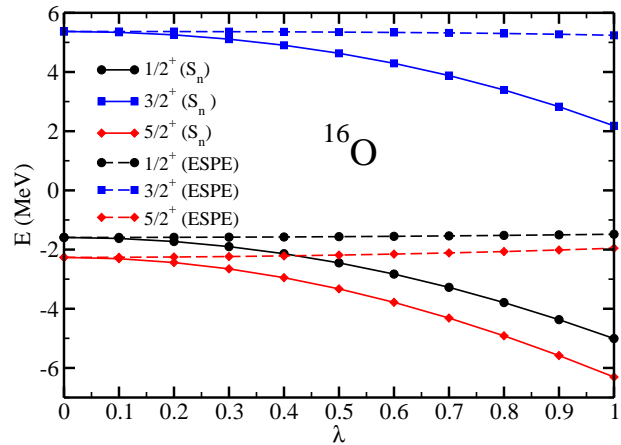}
\caption{(Color online) One-neutron addition energies $E^{+}_\mu$ on $^{16}$O ground-state and corresponding ESPEs $e^{\text{cent}}_p$ as a function of the residual interaction strength.}
\label{switching_on1}
\end{figure}

As a first example, Figure~\ref{switching_on1} displays, as a function of the residual interaction strength, one-neutron separation energies between $^{16}$O ground-state and low-lying states in $^{17}$O along with corresponding ESPEs. Plotted separation energies correspond to the (main) lowest peak in the additional sector of the SDD for each $J^{\pi}$ symmetry block. As expected, one-neutron separation energies and ESPEs are equal in the uncorrelated limit ($\lambda=0$) and are nothing but HF single-particle energies, i.e. Koopmans' theorem is fulfilled. Turning on correlations, two important features manifest themselves. First, the SDD is fragmented such that the separation energy of the state carrying the largest strength for a given $J^{\pi}$ goes down significantly. Second, correlations only slightly impact centroid energies that keep a strong memory of HF single-particle energies. Thus, although ESPEs are not independent of correlations, the latter are essentially screened out as discussed earlier. Eventually, one-nucleon separation energies and corresponding ESPEs can differ by several MeVs. This clearly points to the fact that separation energies should not be
identified as ESPEs and vice versa.

\begin{figure}
\includegraphics[width=0.45\textwidth,clip=]{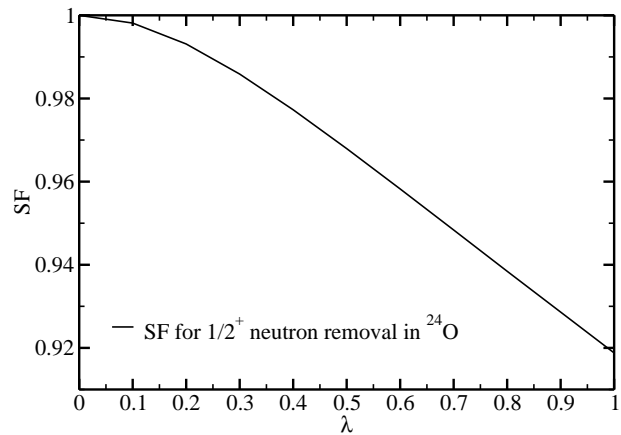}
\caption{(Color online) Removal spectroscopic factor $SF^{-}_{1/2^+}$ of the lowest $1/2^+$ state in $^{23}$O as a function of the residual interaction strength.}
\label{switching_on2}
\end{figure}

\begin{figure}
\includegraphics[width=0.45\textwidth,clip=]{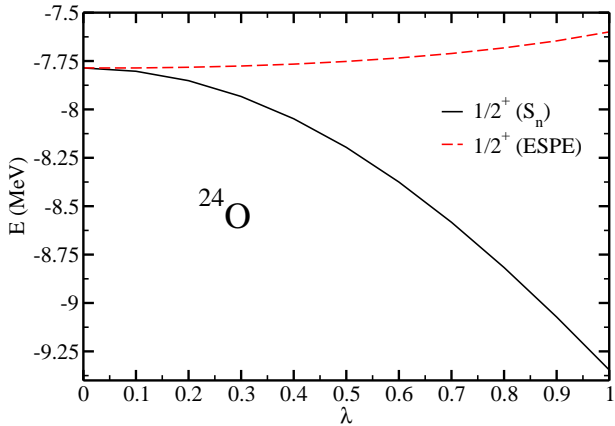}
\caption{(Color online) Same as Figure~\ref{switching_on2} for one-neutron removal energy $E^{-}_{1/2^+}$ and  ESPE $e^{\text{cent}}_{2s_{1/2}}$.}
\label{switching_on3}
\end{figure}

Let us now include more detail by focusing on the lowest $1/2^+$ state in $^{23}$O, which can be accessed by removing a neutron from the ground state of $^{24}$O. Figure~\ref{switching_on2} displays the corresponding spectroscopic factor $SF^{-}_{1/2^+}$ as a function of the residual interaction strength. While $SF^{-}_{1/2^+}=1$ for $\lambda=0$ as expected, it decreases gently as the residual interaction is switched on to reach a value of 0.92 in the fully correlated case. Using it as a (scale-dependent) analysis tool, such a spectroscopic factor tells us that the lowest $1/2^+$ state keeps a well-pronounced single-particle character even in the fully correlated limit.


Figure~\ref{switching_on3} shows corresponding neutron separation energy and ESPE as a function of $\lambda$. In the uncorrelated limit, i.e. $\lambda=0$, the separation energy and ESPE coincide, while for larger $\lambda$ they start to deviate. While $e^{\text{cent}}_{2\,1/2+}$ is only slightly influenced by correlations, $E^{-}_{1/2+}$ dives significantly as $\lambda$ increases from 0 to 1. Eventually, correlations add about 1.5\,MeV to the separation energy, such that it differs from $e^{\text{cent}}_{2\,1/2+}$ by 1.7\,MeV for $\lambda=1$. Even though the $1/2^+$ state retains to a large extent its single-particle nature, its energy is strongly impacted by correlations and does not provide clean information about the effective single-particle shell structure.

\subsection{Systematics in oxygen and calcium isotopes}
\label{results1a}

We now discuss the evolution and trends of low-lying one-neutron addition and removal energies together with ESPEs in doubly closed-shell oxygen and calcium isotopes. The present calculations are performed with the specific cutoff value $\Lambda = 2.4$ fm$^{-1}$. As a result, the neutron drip line in oxygen and calcium isotopes is wrongly predicted to be located beyond $^{28}$O and $^{60}$Ca, respectively. Three-body forces seem to be mandatory to correctly reproduce the drip line location at $^{24}$O~\cite{Otsuka:2009cs} for oxygen isotopes. It remains to be seen in which way forces of higher rank modify qualitatively or quantitatively the conclusions of the present investigation.

\begin{figure}
\includegraphics[width=0.45\textwidth,clip=]{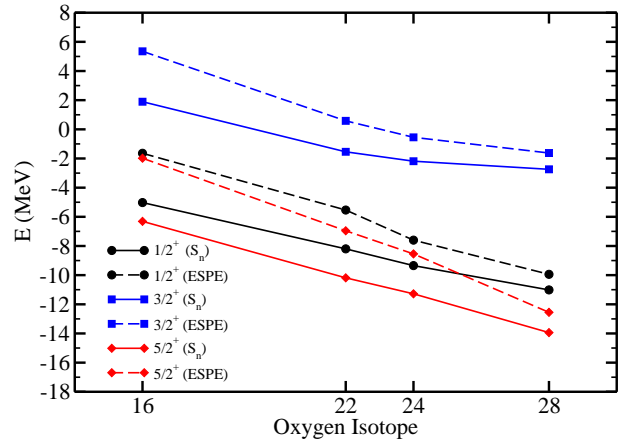}
\caption{(Color online) Evolution of selected one-neutron separation energies $E^{+}_\mu$ and corresponding ESPEs $e^{\text{cent}}_p$ from $^{16}$O to $^{28}$O.}
\label{systematicoxygen}
\end{figure}

\begin{figure}
\includegraphics[width=0.45\textwidth,clip=]{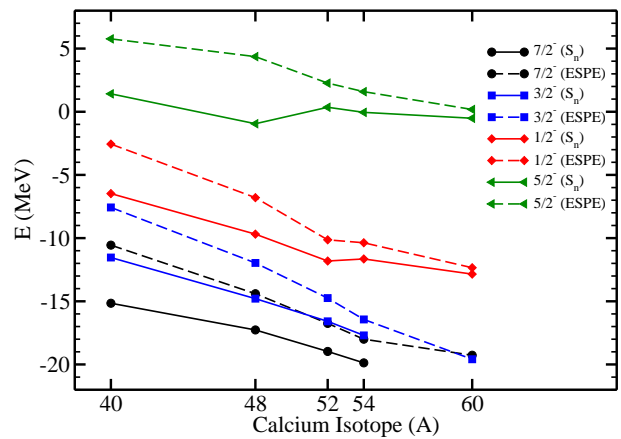}
\caption{(Color online) Same as Figure~\ref{systematicoxygen} from $^{40}$Ca to $^{60}$Ca.}
\label{systematiccalcium}
\end{figure}

Figures~\ref{systematicoxygen} and~\ref{systematiccalcium} show that separation energies $(E^+_\mu ,E^-_\nu)$ systematically and significantly differ from corresponding centroid energies. As for the energetics, these results illustrate that ab-initio approaches describe doubly magic nuclei such as $^{16,24}$O and $^{40,48}$Ca as strongly correlated systems. Most importantly, how much separation energies differ from centroid energies significantly depends on the nucleus/state, in a way that cannot easily be traced back to one specific feature. Consequently, opening or closing of shell gaps in the separation energy spectrum are not in one-to-one relationship with those emerging in the ESPE spectrum. One does observe that $^{16}$O and $^{40}$Ca display the strongest correlations of all, which may be related to their N$=$Z character. Tracing the isospin dependence of correlations in oxygen and calcium isotopic chains, there seems to be a systematic trend with increasing asymmetry N-Z. In both oxygen and calcium chains, correlations become less important for the neutrons close to the Fermi surface when increasing isospin asymmetry N-Z.
This trend is consistent with Ref.~\cite{Jensen2011b}, where it was found that the spectroscopic factor for
removing a neutron close to the Fermi surface increases with increasing isospin asymmetry and is close to one for $^{28}$O,
while the spectroscopic factor for removing the outermost protons is largely quenched
with increasing isospin asymmetry.

One can conclude from Figs.~\ref{systematicoxygen} and~\ref{systematiccalcium} that inferring one-nucleon separation energies from ESPEs is not straightforward, even in doubly closed-shell nuclei. One should thus simply not use one for the other.



\subsection{Effective shell model}
\label{SM}

Equation~\ref{HFfield} was obtained following an ab-initio strategy, i.e. considering all nucleons as active and interacting via realistic 2N and 3N interactions in a large enough single-particle Hilbert space. In the traditional effective shell-model, however, the equivalent of Eq.~(\ref{HFfield}) is derived from an effective Hamiltonian defined for $n^{\text{val}}$ active nucleons in a restricted valence space above a closed core composed of $n^{\text{core}}$ nucleons and below an excluded space. 

In such a context, ESPEs are "anchored" on (experimental) one-nucleon addition energies to the core nucleus, i.e. $e^{\text{core}}_{p} \equiv E^{+}_\mu \, \delta_{pk}$. As seen in previous sections, this constitutes in fact a bad approximation when taking an ab-initio perspective. However, and as confirmed by the present investigation, low-lying states obtained by adding (removing) one nucleon to (from) a doubly closed shell nucleus do possess a well-defined single-particle character. The reason why the corresponding ESPE differs significantly from the separation energy is due to the fact that the former collects small strength rejected to rather high missing energies. Eventually, the fact that low-lying states carry most of the strength makes them good candidates to represent {\it quasi} single-particle degrees of freedom. Such a textbook result constitutes the basic justification for the effective shell model that omits the high-lying fragmented strength and recollects the full strength into low-lying states of the core+one-nucleon system.

\subsection{Using a fixed single-particle basis}
\label{results2}

Computing ESPEs in an approximate fashion generates a model dependence that may compromise their usefulness. It is, for example, customary to use uncorrelated occupations of single-particle states in place of the correlated one-body density matrix in Eq.~(\ref{HFfield}) and/or to define ESPEs as the diagonal matrix elements of the centroid field $h^{\text{cent}}$ in an a priori chosen single-particle basis, e.g. a harmonic oscillator basis, rather than as its eigenvalues. The latter approximation is formally questionable as it provides a basis-dependent definition of ESPEs, the quality of which depends on the realistic character of the chosen basis. In practice, the quantitative impact of such an approximate scheme depends on the situation.

\begin{figure}[h]
\includegraphics[width=0.45\textwidth,clip=]{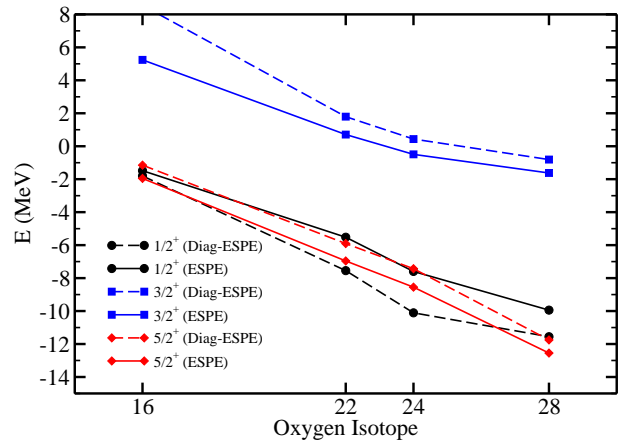}
\caption{(Color online) ESPEs $e^{\text{cent}}_p$ compared to diagonal matrix elements of the centroid field $h^{\text{cent}}_{qq}$ in the underlying HF basis. Results are displayed from $^{16}$O to $^{28}$O.}
\label{systematicoxygen2}
\end{figure}

Figure~\ref{systematicoxygen2} compares in oxygen isotopes properly computed ESPEs with diagonal matrix elements of $h^{\text{cent}}$ in the HF basis used in the calculation\footnote{A more drastic approximation not shown here consists in using diagonal matrix elements in a harmonic oscillator basis. This is the choice usually made within the frame of the interacting shell model. In practice the model space is usually small enough to contain only one state per symmetry block, e.g. ($l,j,m$) block in spherical symmetry. In such a case the arbitrarily chosen basis is necessarily the eigenbasis of the centroid field, underpinning the strong impact of using a severely restricted model space.}. As can be inferred from the comparison with Fig.~\ref{systematicoxygen}, the approximation induces errors on ESPEs that are of the same order as their difference with one-nucleon separation energies and that are, in some cases, significant relative to their absolute values. The error depends both on the state and on the system, i.e. it might go in opposite directions depending on the state and/or the nucleus under consideration.

Interestingly, there exists cases for which the ordering of approximate ESPEs at the Fermi level is inverted compared to full-fledged ones, e.g. for $2s_{1/2}$ and $1d_{5/2}$ levels in $^{22}$O. Knowing that full-blown ESPEs reproduce the ordering of one-neutron separation energies across the whole set of oxygen and calcium isotopes, such an inversion is of noticeable importance. The inversion seen in $^{22}$O is consistent with Fig. 2a of Ref.~\cite{Otsuka:2009cs} where ESPEs were computed as diagonal matrix elements of $h^{\text{cent}}$ in an a priori chosen harmonic oscillator basis. The fact that $2s_{1/2}$ and $1d_{5/2}$ levels are actually not inverted in Fig. 2a of Ref.~\cite{Otsuka:2009cs} is simply due to the fact that ESPEs are anchored on {\it empirical} values in $^{17}$O. Correcting mentally for such a fact, one recovers the level inversion seen for "Diag-ESPE" in Fig.~\ref{systematicoxygen2} of the present paper.


\subsection{Resolution scale dependence}
\label{results2a}

A more fundamental model dependence of ESPEs that remains even when computing them as eigenvalues of the centroid matrix relates to the resolution scale characterizing the Hamiltonian. We start from a Chiral Hamiltonian built with a cutoff $\Lambda_{\chi}$ (e.g. 500\,MeV here) up to a given order (e.g. N$^{3}$LO here). This in itself carries a truncation error with respect to using the complete EFT Lagrangian. Still, this constitutes our reference Hamiltonian, which at N$^{3}$LO contains both 2N and 3N interactions. In a second step, the resolution scale of the Hamiltonian is lowered to a value $\Lambda$ through a renormalization group transformation, defining in this way $H(\Lambda) \equiv H_{\text{low-k}}$. Doing so softens the interactions and induces multi-body forces, e.g. 3N interactions are induced from the original 2N one, while preserving the original truncation error. As $\Lambda$ is lowered, true observables remain the same as with the original Chiral Hamiltonian as long as induced interactions are kept in the calculation and the many-body problem is solved exactly. Contrarily, even in such conditions non-observables quantities such as ESPEs are modified when changing $\Lambda$. This constitutes the {\it intrinsic} scale dependence of ESPEs discussed in Sec.~\ref{sec_scale_dependence} and that we presently wish to characterize. Of course, whenever induced interactions are discarded and/or the many-problem is not solved exactly, both observable and non-observable quantities acquire an additional {\it artificial} dependence on $\Lambda$. 

\begin{figure}
\includegraphics[width=0.45\textwidth,clip=]{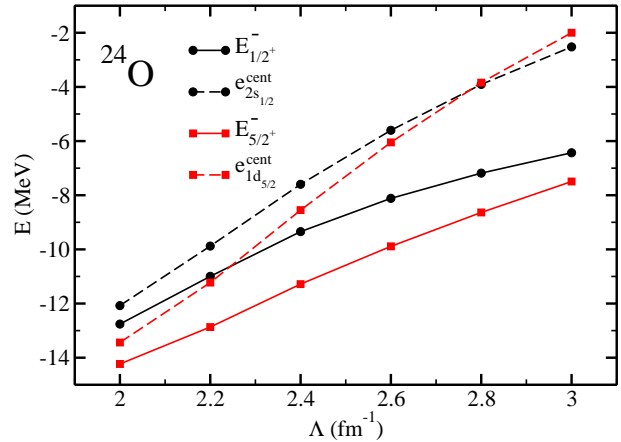}
\caption{(Color online) Neutron ESPEs and removal energies in $^{24}$O for $J^\pi = 1/2^+, 5/2^+$ and $\Lambda \in [2.0,3.0]$ fm$^{-1}$.}
\label{ox24cutoffdep}
\end{figure}

As original and induced three-body forces, as well as clusters beyond singles and doubles, are discarded in the present calculation, ESPEs display the two sources of $\Lambda$ dependence. In order to extract the intrinsic one, one must first pin down the artificial scale dependence to subtract it eventually. By definition, the latter can be accessed by focusing on true observables. Figure~\ref{ox24cutoffdep} displays one-neutron removal energies with  $J^\pi = 1/2^+, 5/2^+$ in $^{24}$O for various values\footnote{We keep the oscillator frequency fixed at $\hbar\omega$ =16\,MeV in present calculations. For large cutoff values, the optimal oscillator frequency is larger, e.g. $\hbar\omega$ =24\,MeV for $\Lambda$ = 2.6\,fm$^{-1}$, such that corresponding values shown in Fig.~\ref{ox24cutoffdep} are not fully converged, e.g. they changed by about 100\,keV for $\Lambda$ = 2.6\,fm$^{-1}$ when using $\hbar\omega$ =24\,MeV. Conclusions of the present section are however not modified by using fully converged values.} of the momentum cutoff $\Lambda$ of the 2N interaction $V_{\text{low-k}}$. Lowering $\Lambda$ from $3.0$ to $2.0$ fm$^{-1}$ changes one-neutron removal energies by about 7 MeV. Eventually, including induced many-body interactions, i.e. three- and possibly four-body forces~\cite{Jurgenson:2009qs,Roth:2011ar}, and including triples will remove such an artificial dependence of one-neutron removal energies on $\Lambda$.

Figure~\ref{ox24cutoffdep} also shows ESPEs $e^{\text{cent}}_{2s_{1/2}}$ and $e^{\text{cent}}_{1d_{5/2}}$ in $^{24}$O. Clearly, they display a significantly larger cutoff variation than corresponding one-neutron removal energies. Such a feature, visible in all isotopes and for all states, is identified with the additional intrinsic scale dependence of ESPEs. Mentally subtracting the cutoff dependence of one-neutron removal energies, one sees that such an intrinsic scale dependence increases with $\Lambda$ as the system becomes less and less perturbative, making ESPEs differ more and more from separation energies. Quantitatively speaking, the intrinsic cutoff dependence of  $e^{\text{cent}}_{2s_{1/2}}$ and $e^{\text{cent}}_{1d_{5/2}}$ amounts to about 6 MeV when varying $\Lambda$ from $2.0$ to $3.0$ fm$^{-1}$, which is obviously significant. More specifically, one notes that both ESPEs do not vary identically across the range of $\Lambda$ values. As a matter of fact, one observes an inversion of the ESPE ordering that is not reflected in one-neutron separation energies. Besides removing (most of) their artificial cutoff dependence, it will be of interest to see how much 3N forces modify the intrinsic scale dependence of ESPEs. It is anyway likely that the latter will remain significant.

The above result demonstrates that ESPEs are not absolute and can be changed significantly by modifying mildly the character of the Hamiltonian, i.e. by varying $\Lambda$ over a rather limited range of values, while keeping true observables invariant. Consequently, extracting {\it the} single-particle shell structure and its evolution, e.g. with isospin, from experimental data is an illusory objective. However, it remains possible to perform a meaningful, i.e. internally consistent, analysis of a set of experimental data by extracting ESPEs through consistent structure and reaction models based on the {\it same} nuclear Hamiltonian. Eventually, conclusions regarding the extracted shell structure, e.g. its evolution with isospin, will however  necessarily remain resolution-scale dependent. In particular, working at large scales is somewhat inappropriate in the sense that the extracted shell structure will not fit with the phenomenological low-energy picture in such a case~\cite{Polls94}.

\section{Conclusions}
\label{conclusions}

The present work discusses, from an ab-initio standpoint, the definition, the meaning, and the usefulness of ESPEs in doubly closed shell medium-mass nuclei. Illustrating the various points with state-of-the-art coupled-cluster calculations, the following conclusions are reached.

\begin{itemize}
\item A meaningful single-particle shell structure fulfilling a minimal set of properties and known limits, such as being independent of the particular single-particle basis one is working with, can be extracted from correlated one-nucleon separation energies and associated spectroscopic amplitudes. Such a definition relates effective single-particle energies (ESPEs) to the so-called centroid eigenvalues introduced by Baranger~\cite{baranger70a}.
\item The corresponding non-interacting problem is governed by the one-body centroid field $h^{\text{cent}}$, which sums the kinetic energy and the {\it energy-independent} part of the irreducible one-nucleon self-energy that naturally arises in self-consistent Green's-function methods.
\item It is customary in low-energy nuclear theory to compute ESPEs in an approximate fashion, e.g. by defining them as diagonal matrix elements of $h^{\text{cent}}$ in an a priori chosen single-particle basis rather than as its eigenvalues. We have illustrated the fact that such approximations are unsafe.
\item Even when fulfilling the required set of minimal properties, ESPEs are not strictly observable as they intrinsically depend on the resolution scale $\Lambda$ of the Hamiltonian, i.e. they change under a unitary transformation of the Hamiltonian while true observables remain invariant.  We have indeed demonstrated that ESPEs vary substantially when modifying mildly the resolution scale, i.e. when scanning a rather limited range of $\Lambda$ values while correcting for the artificial dependence due to the omission of induced short-range many-forces. Such a result demonstrates that the objective of extracting a {\it unique} single-nucleon shell structure from correlated observables, e.g. pinning down {\it the} nuclear shell evolution from experimental data, is intrinsically illusory. Still, it is possible to perform a consistent analysis of experimental data and extract a meaningful shell structure. To do so, one must use consistent structure and reaction models based on the {\it same} nuclear Hamiltonian. Eventually, conclusions regarding the extracted shell structure will anyway remain resolution-scale dependent; i.e. two practitioners using (the same) consistent method but starting from different, though unitarily equivalent, Hamiltonians will extract different single-nucleon shell structures from identical observables, e.g. spectra and cross sections. This constitutes a puzzling but important result that sheds a new light on how one should look at the single-particle shell structure. In particular, working at large scales is somewhat inappropriate in the sense that the extracted shell structure will not fit with the phenomenological low-energy picture in such a case.
\item Extracting an effective single-particle shell structure is often done for interpretation/analysis purposes and sometimes done to infer the behaviour of actual observables that are believed to be strongly correlated to patterns in the ESPE spectrum. In the present paper, we have focused on one-nucleon separation energies to low-lying states around good closed-shell nuclei. The conclusion is that correlations are too strong, even with low-scale interactions, for such separation energies to be in quantitative (sometimes qualitative) correspondence with effective single-particle energies around the Fermi energy. This is true even for states that retain a strong single-particle character, i.e. states carrying spectroscopic factors close to one. In a forthcoming study, the same type of analysis will be performed in connection with the energy of the $2^+$ excited state in good closed shell nuclei, i.e. we will study how much such excitation energies correlate with the Fermi gap in the ESPE spectrum.
\item The present study was conducted on the basis of two-nucleon interactions only. It remains to be seen to which extent forces of higher rank modify our conclusions. At the price of computing ESPEs correctly, i.e. as eigenvalues of the centroid matrix rather than as its diagonal matrix elements in an a priori given (harmonic oscillator) basis, the authors of Refs.~\cite{Otsuka:2009cs,Holt:2010yb} could easily repeat the present analysis within the frame of the shell model and characterize the impact of three-nucleon forces in a systematic way.
\item In the present work, ESPEs were defined based on the hypothesis that eigenstates of the nuclear Hamiltonian are also eigenstates of the particle number operator. Ab-initio calculations of open-shell nuclei are currently being developed on the basis of many-body methods breaking particle-number symmetry, i.e. using methods formulated over Fock space rather than over the Hilbert space associated with a definite number of particles. This is the case of the so-called self-consistent Gorkov-Green's function theory~\cite{soma11a}. Extending the definition of ESPEs accordingly~\cite{soma11a}, one will be able to address properties of ESPEs in open-shell nuclei and conclude on their relevance in such a context.
\end{itemize}

\begin{acknowledgments}
The authors are grateful to C. Barbieri for extensive discussions regarding the meaning of quasi-particle degrees of freedom and the validity of the interacting shell model. The authors also thank J. D. Holt, T. Papenbrock, J. Sadoudi and V. Som\`a for useful discussions and suggestions. T. D. thanks B. A. Brown, A. Obertelli and A. Signoracci for useful discussions as well as V. Lapoux for discussions that initiated the present work. This work was supported by the Office of Nuclear Physics, U. S. Department of Energy (Oak Ridge National Laboratory); and No. DE-FC02-07ER41457 (UNEDF SciDAC). G. H. acknowledges support from Espace de Structure Nucl\'eaire Th\'eorique (ESNT). This work was supported partially through FUSTIPEN (French-U.S. Theory Institute for Physics with Exotic Nuclei) under DOE grant number DE-FG02-10ER41700.
\end{acknowledgments}

\begin{appendix}

\section{Useful identities}
\label{identities}

Using Wick's theorem, one can demonstrate the following identities
\begin{widetext}
\begin{eqnarray}
\{[\ad{p},\ac{r} \ad{s}],\ac{q}\} &=& + \delta_{pr} \, \delta_{qs} \,\,\,,  \nonumber \\
\{[\ad{p}, \ac{r} \ac{s} \ad{t} \ad{v}],\ac{q}\} &=&
+ \delta_{pr} \, \delta_{qv} \, \ac{s} \ad{t} - \delta_{pr} \, \delta_{qt} \, \ac{s} \ad{v} \nonumber \\
&&
- \delta_{ps} \, \delta_{qv} \, \ac{r} \ad{t} + \delta_{ps} \, \delta_{qt} \, \ac{r} \ad{v}  \,\,\,, \nonumber \\
\{[\ad{p}, \ac{r} \ac{s} \ac{t} \ad{w} \ad{v} \ad{u}],\ac{q}\} &=& + \delta_{pr} \, \delta_{qu} \, \ac{s} \ac{t} \ad{w} \ad{v} - \delta_{pr} \, \delta_{qv} \, \ac{s} \ac{t} \ad{w} \ad{u} + \delta_{pr} \, \delta_{qw} \, \ac{s} \ac{t} \ad{v} \ad{u} \nonumber \\
&&
-\delta_{ps} \, \delta_{qu} \, \ac{r} \ac{t} \ad{w} \ad{v} + \delta_{ps} \, \delta_{qv} \, \ac{r} \ac{t} \ad{w} \ad{u} - \delta_{ps} \, \delta_{qw} \, \ac{r} \ac{t} \ad{v} \ad{u}\nonumber \\
&&
+\delta_{pt} \, \delta_{qu} \, \ac{r} \ac{s} \ad{w} \ad{v} - \delta_{pt} \, \delta_{qv} \, \ac{r} \ac{s} \ad{w} \ad{u} + \delta_{pt} \, \delta_{qw} \, \ac{r} \ac{s} \ad{v} \ad{u} \,\,\,. \nonumber
\end{eqnarray}
\end{widetext}

\end{appendix}

\bibliography{espe}

\end{document}